\documentclass{aa}  
\usepackage{txfonts}
\usepackage[normalem]{ulem}

\usepackage[dvipsnames,svgnames,table]{xcolor}
\usepackage[utf8]{inputenc} 
\usepackage[T1]{fontenc}    
\usepackage{hyperref}       
\usepackage{url}            
\usepackage{booktabs}       
\usepackage{amsfonts}       
\usepackage{nicefrac}       
\usepackage{microtype}      
\usepackage{footmisc}
\usepackage{graphicx}
\usepackage{natbib}

\usepackage{textgreek}
\usepackage{tabu}
\usepackage{makecell}
\usepackage{listings}
\usepackage{amsmath}
\usepackage{color}
\usepackage{multirow}
\usepackage{multiobjective}
\usepackage[ruled,vlined]{algorithm2e}

\usepackage{xspace}
\usepackage{tikz}
\usetikzlibrary{positioning, arrows, shapes, calc}
\usetikzlibrary{shapes.multipart}

\begin{document} 
   \title{Comparing phylogenetic trees of stars with spectral graph distances}
   \author{Theosamuele Signor          \inst{1,2}\thanks{\email{theosamuele.signor@mail.udp.cl}}
          \and
          Matias Neto\inst{2}
          \and
          Paula Jofré\inst{1}
          \and
          Sara Vitali\inst{3,4}
          \and
          Xia Hua\inst{5}
          \and
          Patricia B. Tissera\inst{6}
          \and
          Claudia Aguilera-Gómez\inst{6}
          \and
          Payel Das\inst{7}\and
          Brian Tapia-Contreras\inst{6,8}
          \and
          Robert M. Yates\inst{9}
          \and
          Luis Martí\inst{2}
          \and
          Nayat Sánchez-Pi\inst{2}
          }
   \institute{Instituto de Estudios Astrofísicos, Facultad de Ingeniería y Ciencias, Universidad Diego Portales, Av. Ejercito 441, Santiago, Chile
         \and
         Inria Chile Research Center, Av. Apoquindo 2827, piso 12, Las Condes, Santiago, Chile
         \and
         Instituto de Astrofísica de Canarias, C. Vía Láctea, s/n, 38205 La Laguna, Santa Cruz de Tenerife, Spain
         \and
         Universidad de La Laguna, Dpto. Astrofísica, Av. Astrofísico Francisco Sánchez, 38206 La Laguna, Santa Cruz de Tenerife, Spain
         \and 
         Mathematical Sciences Institute, Australian National University; Acton, ACT 0200, Australia
         \and
         Instituto de Astrofísica, Pontificia Universidad Católica de Chile, Av. Vicuña Mackenna 4860, Santiago, Chile
         \and 
         School of Mathematics and Physics, University of Surrey, Guildford, Surrey, GU2 7XH, UK
         \and
         Centro de Astro-ingeniería, Pontificia Universidad Católica de Chile, Av. Vicuña Mackenna 4860, Santiago, Chile
         \and 
         Centre for Astrophysics Research, University of Hertfordshire, Hatfield, AL10 9AB, UK
             }
   \date{Received 24 July 2026; accepted 20 September 2026}
 
  \abstract
   {Phylogenetic trees offer a hierarchical representation of chemical similarities among stars, from which their evolutionary histories can, in principle, be reconstructed.
   Simulation-based work suggests that phylogenetic trees built under different chemical evolution scenarios carry distinguishable structural imprints, but whether existing methods can detect this signal in practice is unclear.
   }
   {We evaluate a range of distance measures, both classical tree-based and graph-theoretical, in their ability to compare such trees, seeking measures that are robust to observational noise and that separate trees built from populations with distinct chemical enrichment histories.
   }
   {We construct phylogenetic trees using the Neighbor-Joining algorithm from chemical abundance vectors of solar twins and of stars associated with the Milky Way (MW) disk, Gaia–Sausage–Enceladus (GSE), and the Sagittarius dwarf spheroidal galaxy (Sgr). We then compare trees using classical tree-based and graph-based measures.
   Two benchmark tests assess: (1) stability under input perturbations within the same evolutionary history, and (2) discriminative power between distinct stellar populations used as proxies for different evolutionary histories.}
  {Classical tree metrics rapidly saturate under perturbations of order the typical abundance uncertainty ($\approx2\sigma$, or $\approx0.05$ dex per element) and fail to distinguish trees when leaf sets differ.
  In contrast, spectral distances vary smoothly with perturbation amplitude and provide a label-invariant framework for comparing trees built from disjoint sets of stars, including populations with distinct chemical evolutionary histories.
  Among the spectral measures evaluated, the Laplacian spectral distance (LSD) shows the most consistent separation across population pairs, while the normalized variant (NLSD) is the weakest due to its reduced sensitivity to branch lengths.
  Applied to MW, GSE, and Sgr stars, spectral measures recover a distance ordering consistent with the known chemical enrichment histories of these populations.
  Beyond distinguishing already well-characterized populations, this framework offers a route to comparing, for example, an observed enrichment history against simulated ones to identify which best reproduces the observed tree structure.
  }
   {}
   \maketitle

\section{Introduction}\label{sec:intro}
Phylogenetic methods have a long history in fields like biology and linguistics \citep{baum2005tree, gray2009language}, where they are used to reconstruct evolutionary relationships between species or languages. In recent years, these techniques have found new applications in astrophysics \citep{jofre2017cosmic, jackson2021using}, leading to the development of galactic phylogenetics. Here, individual stars are treated analogously to biological taxa, and their chemical abundance patterns, enriched by earlier generations of stars, act as heritable features encoding information about their formation environments \citep{Freeman2002newgalaxy}. These methods have been successfully applied to observed datasets, revealing hierarchical structures that might trace different Galactic enrichment pathways \citep{jofre2025studying}.

Thanks to large spectroscopic surveys such as GALAH \citep{desilva2015galah} and APOGEE \citep{majewski2017apache}, it is now possible to obtain precise chemical abundances for millions of stars across the Milky Way. When combined with astrometric data from Gaia \citep{Vallenari2023gaiadr3}, which provides accurate positions and proper motions, these datasets provide the basis for detailed studies of the Galaxy’s structure, kinematics, and chemical enrichment history.

These efforts are often pursued through strong chemical tagging \citep{Freeman2002newgalaxy}, which treats stellar elemental abundances as high-dimensional fingerprints for identifying co-natal stars. In practice, this involves applying clustering algorithms, such as k-means \citep{lloyd1982least}, DBSCAN \citep{ester1996density}, or HDBSCAN \citep{campello2013density}, in chemical space to group stars with similar abundance patterns. 
However, different birth environments can yield similar chemical signatures due to overlapping enrichment pathways, stellar migration, and mixing processes \citep{edvardsson1993chemical, bird2012radial, magrini2023mapping}, as well as intrinsic abundance scatter arising from chemical inhomogeneities in the natal gas cloud. Observational uncertainties further blur group boundaries \citep{blancocuaresma2015testing, bovy2016chemical}. As a result, recent studies have shown that strong chemical tagging alone may be insufficient to uniquely recover birth associations (e.g. \citealt{barth2025trials, Signor2024baseline, spina2022chemical, casamiquela2021impossibility}). 
This motivates alternative approaches that reconstruct hierarchical, continuous relationships between stars instead of assigning them to discrete groups.

Galactic phylogenetics implements this: it seeks to reconstruct evolutionary relationships among stars, organizing them into trees based on chemical similarity. In this framework, a phylogenetic tree can be interpreted as encoding shared chemical ancestry: both its branching pattern (topology) and branch lengths reflect patterns of chemical divergence.
Comparing phylogenetic trees is then conceptually distinct from comparing chemical abundances directly: rather than how far apart two stars sit in abundance space, it asks whether two populations share the same (or similar) branching organization. Two populations may differ in individual abundances yet share similar large-scale tree structure.

A central question in galactic phylogenetics is whether phylogenetic trees encode structural signatures of the underlying evolutionary history, and whether this signal can be reliably detected in practice.
Recent simulation-based work has begun to address the first part of this question.
Analytical chemical evolution models show that trees can separate populations evolving under distinct star formation conditions \citep{canales2026disentangling}.
Hydrodynamical simulations of isolated disc galaxies demonstrate that tree morphology encodes enrichment 
history across galactic regions \citep{deBritoSilva2024evolutionary, 
tapia2026reconstructing}.
These results support the hypothesis that chemical evolutionary histories leave recoverable structural imprints in phylogenetic trees.

The second part of the question, whether existing comparison methods are adequate to detect this signal in practice, remains open. Existing approaches typically characterise a tree through a single structural summary. Tree balance, quantified for instance by the Corrected Colless index (\citealt{heard1992patterns, kirxpatrick1993searching,fischer2023tree}), distinguishes hierarchical, bursty enrichment from more symmetric, steady star formation, and separates populations robustly down to modest sample sizes \citep{tapia2026reconstructing}. The cumulative branch length, or distance from the root to the tips, traces the intensity and duration of star formation, with more actively star-forming regions producing longer trees \citep{deBritoSilva2024evolutionary}. The shape of individual branch paths has likewise been linked to enrichment history, though so far only qualitatively \citep{canales2026disentangling}. 
Because these are scalar summaries, they can be compared across trees built from different stellar populations, but each captures only a single aspect of tree structure and discards the rest of the topology.

Classical tree distance measures, such as the Robinson–Foulds distance (RF; \citealt{RobinsonFoulds1981}), have the opposite problem. They compare trees through their full structure rather than a single summary. Many, however, ignore branch lengths, despite their relevance for continuous chemical enrichment and observational uncertainty. All require identical leaf sets, preventing comparisons between trees constructed from different stellar populations or disjoint samples.
None of these approaches combine the two properties needed here: sensitivity to the full tree structure and the ability to compare trees with different leaf sets. 

We therefore ask whether tree-comparison measures can distinguish the variability expected between trees associated with the same evolutionary history from differences between trees associated with different populations.
Answering this requires distance measures\footnote{Distance metrics are often used for this purpose. In our case, we consider distances that do not necessarily satisfy the formal properties of a metric, and therefore use the broader term distance measures.} that tolerate local rearrangements induced by noise or sampling but remain sensitive to differences associated with genuinely distinct evolutionary scenarios.

This challenge is amplified by the nature of stellar phylogenies themselves. 
Stellar trees are constructed from individual stars described by a relatively small number of continuous chemical abundances affected by measurement uncertainties. The number of stars can also be large, making visual inspection impractical. Moreover, small perturbations in abundance space may lead to substantial topological rearrangements even when the underlying evolutionary history remains unchanged.
This sensitivity has been demonstrated empirically: trees built from chemically similar stars can change considerably under resampling of the abundances within their uncertainties  \citep{walsen2024assembling}, and the dissimilarity between perturbed trees grows steadily with the abundance uncertainty \citep{deBritoSilva2024evolutionary}.

The limitations of existing measures, together with the instability of stellar trees, motivate a graph-theoretical perspective. A phylogenetic tree is fundamentally a weighted graph (a set of nodes connected by edges, each edge carrying a numerical weight, here related to evolutionary distance), and biologists have drawn on graph-theoretic tools for tree comparison and characterization (see, e.g. \citealt{wilberg2015what,lloyd2016estimating,lewitus2016characterizing}). Rather than comparing trees through tree-specific operations, we compare their graph representations through spectral descriptors: quantities derived from the eigenvalues of the graph, which are invariant to node labeling. That is, they don't depend on which star is assigned to which node, only on the graph's structure. Within this framework, trees can be compared even when built from different sets of stars.
Specifically, we:
\begin{enumerate}
  \item Represent each phylogenetic tree as a weighted graph, where edge weights encode evolutionary distance;
  \item Characterize these graphs using spectral descriptors, which are invariant to node labeling;
  \item Evaluate classical tree and graph-based measures in their ability to distinguish trees generated under shared versus distinct chemical evolution scenarios, using distinct stellar populations as proxies for the latter.
\end{enumerate}
We do not focus on tree construction itself. We assume trees are built, using the Neighbor-Joining (NJ) algorithm \citep{saitou1987neighbor}, which has been the tree-construction method used in galactic phylogenetic studies to date, and focus exclusively on their comparison.

This paper is organized as follows. In Section~\ref{sec:methods}, we describe the mathematical framework for graph representation and distance measures. Section~\ref{sec:data} describes the data used. Section~\ref{sec:results} presents two benchmark experiments that assess the sensitivity and discriminative power of each measure. Finally, in Section~\ref{sec:conclusions}, we discuss the implications of our findings for tree comparison, data representation, and future directions in galactic phylogenetics.

\section{Methods}\label{sec:methods}
We describe the methodology used to compare phylogenetic trees constructed from stellar chemical abundance data.  
Our approach consists of four main stages: constructing trees from chemical abundance vectors (Sect.~\ref{sec:trees}); applying classical tree distance metrics to quantify differences (Sect.~\ref{sec:tree_metrics}); representing trees as graphs (Sect.~\ref{sec:graphs}), and computing graph-based measures (Sect.~\ref{sec:graphs_metrics}).
Section~\ref{sec:benchmark} then outlines the benchmark experiments designed to evaluate the behavior of these measures.

\subsection{Phylogenetic trees}\label{sec:trees}
A phylogenetic tree is a representation of evolutionary relationships among a set of objects. In the context of galactic phylogenetics, the objects are individual stars and the inferred relationships are interpreted as reflecting a shared chemical evolutionary history. This interpretation relies on a central assumption of Galactic archaeology that stellar chemical abundance patterns retain information about the interstellar medium from which stars formed and can therefore be used to trace past enrichment processes.

Under this framework, each leaf of the tree corresponds to a star, while internal nodes represent inferred ancestral chemical states of the interstellar medium. The paths connecting leaves encode chemical divergence, with shorter paths corresponding to more chemically similar stars. A phylogenetic tree therefore encodes both topology (how stars are grouped and branched) and branch lengths, which quantify the degree of evolutionary divergence between objects.
Phylogeny is not created by the tree-building algorithm itself, but emerges from the information content of the abundance data; the algorithm serves only to organize this information into a tree representation.
 
We construct phylogenetic trees from stellar abundance vectors using the Neighbor-Joining (NJ) algorithm \citep{saitou1987neighbor}, applied to a pairwise Manhattan chemical distance matrix computed in chemical abundance space. We refer to these pairwise distances, and the branch lengths they induce in the tree, as evolutionary distances.
For more details we refer to seminal works such as \cite{felsenstein2004inferring, yang2014molecular}.
Since we focus exclusively on tree comparison rather than tree construction, we treat the NJ tree as a reasonable proxy for the chemical relationships among stars, without optimising the choice of tree-building algorithm.

\subsection{Tree-to-tree distance measures}\label{sec:tree_metrics}
To quantify the similarity between two trees, several distance measures have been proposed, each emphasizing different aspects of their topology or branch-length structure.
In what follows, we briefly describe the main families of distances considered in this work; their mathematical definitions are provided in Appendix~\ref{app:tree_metrics}. 

The most commonly used is the Robinson–Foulds distance (RF, \citealt{RobinsonFoulds1981}), a purely topological measure that counts the number of differing splits between two trees. Because it depends only on topology, branch lengths are ignored and all split mismatches contribute equally to the distance. The Weighted Robinson–Foulds distance (WRF) extends RF by incorporating branch lengths into the comparison.

More flexible variants aim to account for partial structural agreement. The Matching Split Distance (MSD; \citealt{Bogdanowicz2012}) pairs similar splits between trees, while the Clustering Information Distance (CID) and Phylogenetic Information Distance (PID; \citealt{Llabres2021}) interpret trees as probabilistic partitions and compare their shared information content. These approaches relax strict topological matching but remain sensitive to differences in tree structure and leaf correspondence.
The Tree Edit Distance (TED) instead measures the minimum cost of transforming one tree into another through node insertions, deletions, and relabelings, where costs are weighted by branch lengths (see Appendix~\ref{app:tree_metrics} for details).

Despite their differences, these measures rely on explicit correspondences between leaf labels, requiring the trees to share the same underlying objects. 
This is natural in their original biological setting, where the set of objects is fixed and the goal is to compare alternative reconstructions of a single true tree; in our problem we instead compare trees built from distinct stellar populations and seek a description of structure that is independent of which objects they contain.
We therefore consider an alternative graph-theoretical formulation, where similarity is assessed through spectral properties of associated matrix operators, providing a label-invariant description of structure.
A qualitative comparison of tree-based and spectral distance measures is provided in Table~\ref{tab:distance_comparison}.
\begin{table*}
\centering
\caption{Qualitative comparison of tree-based and spectral distance measures.}
\begin{tabular}{lccc}
\hline\hline
Measure & Weight sensitivity & Structural scale & Requires leaf correspondence \\
\hline
Robinson–Foulds Distance  & None              & Local (topological)          & Yes \\
Weighted Robinson–Foulds Distance & High              & Local (topological + metric) & Yes \\
Matching Split Distance  & Moderate          & Intermediate                 & Yes \\
Clustering Information Distance  & Moderate          & Global (information-based)   & Yes \\
Phylogenetic Information Distance  & Moderate          & Global (information-based)   & Yes \\
Tree Edit Distance  & High\tablefootmark{(a)}    & Global (edit-based)          & Yes \\
\hline
Adjacency Spectral Distance  & High              & Local / fine-scale           & No  \\
Laplacian Spectral Distance  & Moderate          & Global                       & No  \\
Normalized Laplacian Spectral Distance & Low (normalized)  & Global (degree-normalized)   & No  \\
\hline
\end{tabular}
\tablefoot{Weight sensitivity refers to whether the measure incorporates branch lengths. Structural scale indicates whether the measure captures local or global structure. Leaf correspondence indicates whether the measure requires both trees to be defined on the same set of labeled leaves.
\tablefoottext{a}{TED weight sensitivity depends on the choice of edit cost function — the rule assigning a numerical cost to each edit operation (node insertion, deletion, or relabeling) used to transform one tree into another (see Appendix \ref{app:tree_metrics}). In this work, insertion and deletion costs equal the branch length of the affected node, and relabeling cost equals the absolute difference in branch lengths between the two nodes, making TED sensitive to metric structure throughout.}}
\label{tab:distance_comparison}
\end{table*}

\subsection{From trees to graphs}\label{sec:graphs}
\begin{figure*}[t]
    \centering
    \includegraphics[width=\linewidth]{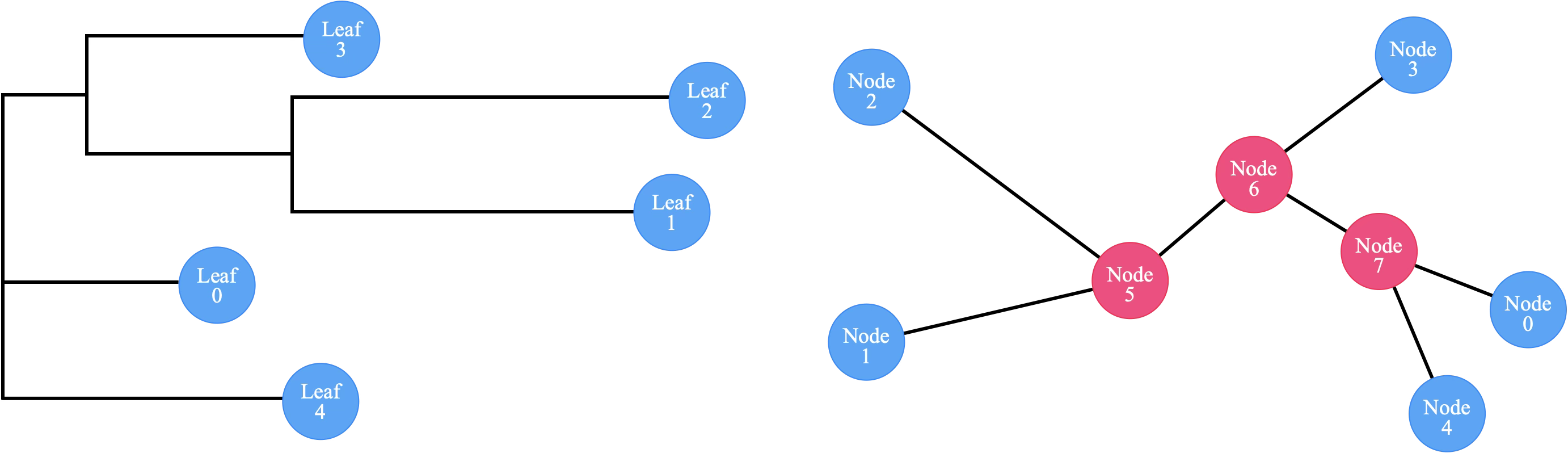}
    \begin{minipage}[t]{0.48\linewidth}
        \centering
        \small
        \texttt{((2:0.84, 1:0.76):0.46, 3:0.49,}\\
        \texttt{(0:0.39, 4:0.62):0.18);}
    \end{minipage}
    \hfill
    \begin{minipage}{0.48\linewidth}
        \centering
        \small
        \[
        \mathbf{A} = \bordermatrix{
  & 0    & 1    & 2    & 3    & 4    & 5    & 6    & 7    \cr
  0   & \cdot & \cdot & \cdot & \cdot & \cdot & \cdot & \cdot & 0.68 \cr
  1   & \cdot & \cdot & \cdot & \cdot & \cdot & 0.47  & \cdot & \cdot \cr
  2   & \cdot & \cdot & \cdot & \cdot & \cdot & 0.43  & \cdot & \cdot \cr
  3   & \cdot & \cdot & \cdot & \cdot & \cdot & \cdot & 0.61  & \cdot \cr
  4   & \cdot & \cdot & \cdot & \cdot & \cdot & \cdot & \cdot & 0.54 \cr
  5   & \cdot & 0.47  & 0.43  & \cdot & \cdot & \cdot & 0.63  & \cdot \cr
  6   & \cdot & \cdot & \cdot & 0.61  & \cdot & 0.63  & \cdot & 0.84 \cr
  7   & 0.68  & \cdot & \cdot & \cdot & 0.54  & \cdot & 0.84  & \cdot \cr
}
        \]
    \end{minipage}
    \caption{
    Representation of a phylogenetic tree (left) as a weighted graph (right) for spectral analysis, shown here for a small illustrative example with five leaves. Leaves (blue terminal nodes) represent individual stellar objects. Internal nodes (red) represent inferred ancestral chemical states, interpreted as proxies for the interstellar medium composition at earlier stages of enrichment.    
    Below each panel, the same tree is expressed in Newick format (left) and as a weighted adjacency matrix (right), where edge weights $w_{ij} = e^{-d_{ij}/\tau}$ (in this case with $\tau=1$) encode the evolutionary distance $d_{ij}$ between connected nodes. Unlike the Newick representation, where internal nodes are implicit, the adjacency matrix treats all nodes as vertices of the graph. 
    For example, in the Newick string, \texttt{(2:0.84, 1:0.76):0.46} indicates that leaves 2 and 1 share an internal node at evolutionary distances of 0.84 and 0.76\,dex respectively, with that internal node itself separated from its parent by 0.46\,dex. In the adjacency matrix, the entry $A_{07} = 0.68$ reflects the weight $w_{07} = e^{-d_{07}}$ assigned to the edge connecting nodes 0 and 7, where $d_{07}$ is their evolutionary distance. Internal nodes 5, 6, and 7 correspond to the three parenthetical groupings in the Newick string, from innermost to outermost.}
    \label{fig:tree-graph}
\end{figure*}
A graph is a collection of nodes connected by edges; here, each node is either a star or an inferred ancestor, and each edge is weighted by how chemically similar the two nodes are.
More formally, a (weighted) graph is a pair $G = (V, E)$ consisting of a set of vertices $V = \{1, \dots, n\}$, and a set of edges $E \subseteq V \times V$. Each edge $(i,j) \in E$ is assigned a positive weight $w_{ij}$ via a weight function 
$W: E \rightarrow \mathbb{R}^+$. We write $i \sim j$ if $(i,j) \in E$.

Graphs can be represented in matrix form. A matrix representation is a $|V| \times |V|$ matrix $X$ where entry $X_{ij}$ encodes some structural property of node pair $(i,j)$.
Different matrix representations capture different aspects of graph structure, and their spectra provide compact descriptors of these properties.
One such representation is the \emph{adjacency matrix} $A \in \mathbb{R}^{n \times n}$:
\[
A_{ij} \equiv 
\begin{cases}
w_{ij}, & \text{if } i \sim j, \\
0, & \text{otherwise}.
\end{cases}
\]
That is, $A_{ij}$ stores the weight of the edge between nodes $i$ and $j$ if one exists, and zero otherwise.
The degree matrix $D$ is diagonal, with $D_{ii} = \sum_{j \sim i} w_{ij}$, and from this we define the combinatorial Laplacian $L = D - A$, and the normalized Laplacian $\hat{L} = D^{-1/2} L D^{-1/2}$ \citep{chung1997spectral}.
The spectrum of the graph is obtained from its chosen matrix representation using the eigendecomposition.

A phylogenetic tree $T = (V, E)$ is a special type of undirected graph that is connected and acyclic, where each edge is assigned a non-negative length representing evolutionary distance or closeness. 
Let $N$ be the number of stellar objects in a sample; then an $N$-tree is a tree with $N$ uniquely labeled terminal nodes, or leaves. Edges connected to leaves are called external, while all others are internal.

All nodes are treated as graph vertices, and edge weights encode connection strength. We define edge weights as:
\begin{equation}
    w_{ij} = e^{-d_{ij}/\tau},
    \label{eq:kernel}
\end{equation}
where $d_{ij}$ is the evolutionary distance between nodes $i$ and $j$ and $\tau>0$ is a scale parameter controlling the sensitivity of the affinity to branch length. 
This weighting function maps distances from the range $[0,\infty)$ to a bounded affinity $w_{ij} \in (0,1]$ and ensures that closer nodes (i.e., more closely related objects) have stronger connections\footnote{
We adopt this form rather than retaining the raw branch lengths ($A_{ij} = d_{ij}$) because bounded weights prevent a few long branches from dominating the spectrum and match the conventional similarity-graph interpretation, under which small Laplacian eigenvalues reflect global connectivity; the alternative $A_{ij} = d_{ij}$ emphasizes divergence rather than similarity and inverts this interpretation.
Under the alternative $A_{ij} = d_{ij}$ convention, the adjacency and Laplacian entries, and hence the spectral quantities, carry the units of the underlying evolutionary scale (dex or dex²); under the adopted $e^{-d_{ij}/\tau}$ weighting they are dimensionless.}. The choice of $\tau$ is data-driven and discussed in Sect.~\ref{sec:results}

The eigenvalues and eigenvectors of matrices like $A$, $L$, or $\hat{L}$ (the graph spectrum) provide a label-invariant embedding of tree structure that can be used for quantitative comparison.
Figure~\ref{fig:tree-graph} illustrates this reinterpretation for a toy five-leaf tree: the phylogenetic tree (left) and its corresponding weighted graph (right) are shown alongside the Newick representation and adjacency matrix.

\subsection{Graph distance measures}\label{sec:graphs_metrics}
Comparing phylogenetic trees can be framed as quantifying structural similarity between graphs.
In this section, we review graph-based spectral measures that can be applied in the context of galactic phylogenetics.
Given the challenges of developing analytical results for random tree ensembles, we evaluate the measures empirically through controlled experiments (see Sect.~\ref{sec:benchmark}; \citealt{monnig2018resistance}).

A generic spectral distance between two graphs \(G_1\) and \(G_2\) is defined as:
\begin{equation}
M_{\text{spec}}(G_1, G_2) = \left( \sum_{i=1}^k \left| \lambda_i^{(1)} - \lambda_i^{(2)} \right|^p \right)^{\frac{1}{p}},
\end{equation}
where $p \ge 1$ is the order of the norm (we adopt $p = 2$ throughout); \(\lambda_i^{(1)}\) and \(\lambda_i^{(2)}\) are the $i$-th eigenvalues of the respective graph matrices, sorted by magnitude following standard spectral conventions (\citealt{chung1997spectral}, descending for adjacency, ascending for Laplacian) and $k\leq \min(|V_1|,|V_2|)$.
Every tree-to-tree comparison in this work is performed between trees with the same number of leaves, and hence the same number of nodes ($|V|=|V_1|=|V_2|$); we therefore use the full spectrum ($k = |V|$) throughout. 

The choice of matrix determines which structural features are emphasized. The Adjacency Spectral Distance (ASD) is more sensitive to variations in edge weights and local connectivity, and thus reflects fine-scale structural differences in the tree, such as variations in branch lengths. The Laplacian Spectral Distance (LSD) emphasizes global structural properties, such as connectivity and large-scale organization.
The normalized Laplacian Spectral Distance (NLSD) applies degree normalization to the graph, reducing sensitivity to node degree and overall edge-weight scale. It is invariant to a uniform rescaling of all edge weights while retaining sensitivity to their relative organization.

\subsection{Benchmark setup}\label{sec:benchmark}
We design two benchmark experiments to evaluate how well different distance measures reflect underlying evolutionary structure. Specifically, we consider two regimes: (i) trees generated from a shared evolutionary history, where distances should remain small under perturbations; and (ii) trees generated from distinct astrophysical formation scenarios, where distances should be large.
This leads to two criteria: consistency, requiring stability under variations within the same evolutionary history, and discriminative power, requiring separation between different formation pathways.

To compare measures with different scales, we rescale each distance $M$ by its maximum value:
\begin{align}
    \widetilde{M}=\frac{M}{\max M},
\end{align}
where the maximum is taken over all realizations in the benchmark.

\subsubsection{Consistency within the same evolutionary history}\label{subsec:sensitivity}
This experiment evaluates the stability of each distance measure under perturbations in the input data.
If a measure captures evolutionary signal rather than noise, distances between reference and perturbed trees should remain small for small perturbations and grow only as the perturbation approaches the point where the underlying chemical similarity is genuinely disrupted.

To evaluate this, we simulate observational uncertainties by perturbing an initial chemical abundance matrix $D \in \mathbb{R}^{N\times m}$ with N stars and m abundance ratios, with Gaussian noise. This independent Gaussian model is an idealized representation of abundance uncertainty. In practice, measurement errors can be correlated across elements and non-Gaussian, which we do not attempt to model here. The resulting perturbed matrix is defined as:
\begin{equation}
    \widetilde{D} = D + \varepsilon,\quad \varepsilon \sim \mathcal{N}(0,\, k^2 \ \vec{\sigma}^2),
    \label{eq:noisy}
\end{equation}
where \(\vec{\sigma}^2\) is the variance associated with the abundance measurements, and \(k\) is a scaling factor controlling the perturbation strength.

For each value of $k \in [0, 5]$, we generate 200 perturbed matrices $\widetilde{D}$, which we interpret as realizations of the same evolutionary history under increasing measurement uncertainty, construct the corresponding phylogenetic trees, and compute their distances to the original reference tree.
This procedure quantifies the stability of each distance measure under increasing perturbations.
Distances should remain low for small k and increase smoothly as perturbations grow, reflecting progressive structural changes without abrupt saturation or insensitivity.

Here the perturbation acts on the abundances while the sample of stars is held fixed, so both trees share the same leaves and all measures, including the classical tree metrics, can be compared directly. The complementary case, where trees are built from different stars drawn from the same population, arises naturally in the discriminative-power benchmark presented in the next section; there different subsamples necessarily have different leaf sets, so only the spectral measures apply.
\subsubsection{Discriminative power across different histories}
\label{sec:hist-difs}
The second criterion evaluates the ability to distinguish between trees derived from different evolutionary histories.
We use chemically distinct stellar populations as proxies for distinct evolutionary histories (see Sect.~\ref{sec:data}).

We compute distributions of pairwise distances from trees sampled within the same population and between different populations. Because within-population trees are built from independent random subsamples, their leaf sets differ as well, so classical leaf-matching metrics cannot be used here either; a successful measure separates the resulting intra- and inter-population distance distributions.

This benchmark evaluates discrimination between populations that are defined independently of the trees themselves, using membership criteria established in Sect.~\ref{sec:data}; it does not attempt unsupervised recovery of population membership, sub-structure, or evolutionary history from tree or graph structure alone. Nor is it designed to establish that spectral tree comparison outperforms a direct comparison of the underlying chemical-abundance distributions; rather, it tests whether a label-invariant representation of tree structure retains the ability to distinguish the stellar populations considered here, and permits comparison between trees built from disjoint sets of stars.

\section{Data}
\label{sec:data}
We use observational stellar abundance datasets from different astrophysical populations.
Every tree-to-tree comparison is performed within a single, internally homogeneous source and trees from different data sources are never compared directly. We draw on five samples, described in detail in the following sections: solar twins from \cite{walsen2024assembling} and \cite{nissen2020high}, and stars associated with the Milky Way (MW), Gaia–Sausage–Enceladus (GSE), and the Sagittarius dwarf spheroidal galaxy (Sgr).

For consistency across datasets, all phylogenetic trees are constructed using the subset of abundances available in every sample:
\[
\begin{aligned}
[\mathrm{Mg}/\mathrm{Fe}],
[\mathrm{Al}/\mathrm{Fe}],
[\mathrm{Si}/\mathrm{Fe}],
[\mathrm{Ca}/\mathrm{Fe}],
[\mathrm{Ti}/\mathrm{Fe}],
[\mathrm{Cr}/\mathrm{Fe}],
[\mathrm{Ni}/\mathrm{Fe}].
\end{aligned}
\]
We adopt $[\mathrm{X}/\mathrm{Fe}]$ abundance ratios, as provided by the datasets used; repeating the analysis with $[\mathrm{X}/\mathrm{H}]$ abundances yields qualitatively and quantitatively similar results. Abundances are used as raw [X/Fe] values without additional standardization; we verify in Appendix~\ref{app:compression} that this choice or the removal of any single chemical dimension does not drive our main results.

\begin{figure}
    \centering
    \includegraphics[width=\columnwidth]{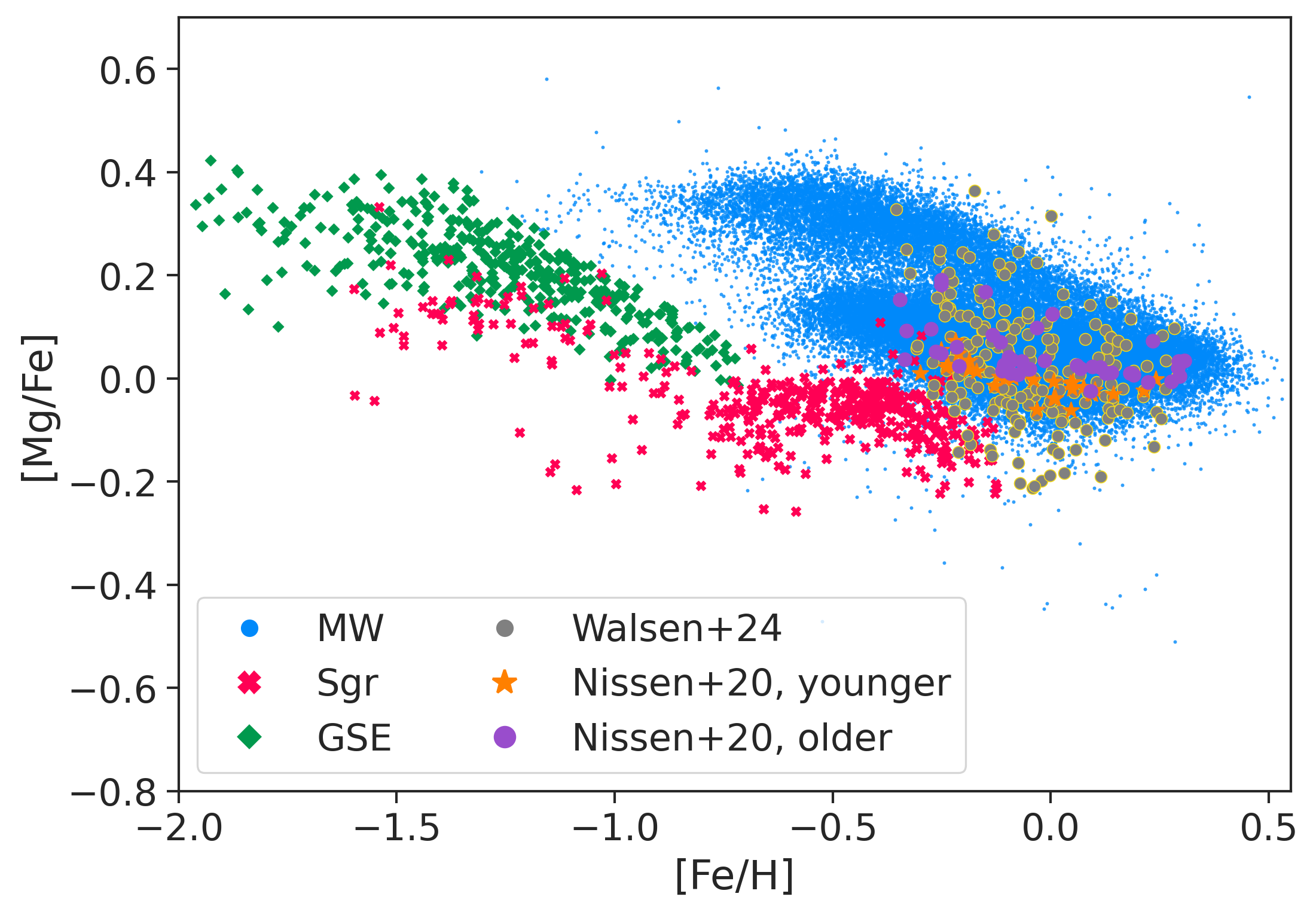}
    \caption{[$\mathrm{Mg}/\mathrm{Fe}$] versus [$\mathrm{Fe}/\mathrm{H}$] for the stellar populations considered in this work.}
    \label{fig:afe_all}
\end{figure}

Figure ~\ref{fig:afe_all} shows the populations in the [$\mathrm{Mg}/\mathrm{Fe}$]--[$\mathrm{Fe}/\mathrm{H}$] plane, where Mg serves as a representative $\mathrm{\alpha}$-element. The [$\mathrm{\alpha}/\mathrm{Fe}$] knee appears at lowest [$\mathrm{Fe}/\mathrm{H}$] for GSE, at intermediate values for Sgr, and highest for the MW disc. 
The position of this knee traces the star-formation efficiency of a population and is widely used in Galactic archaeology to distinguish chemically the histories of accreted and in-situ populations \citep[e.g.][]{matteucci2012chemical}. 
The Nissen and Walsen solar-twin samples sit near solar values.

\begin{figure*}
\centering
\includegraphics[width=\linewidth]{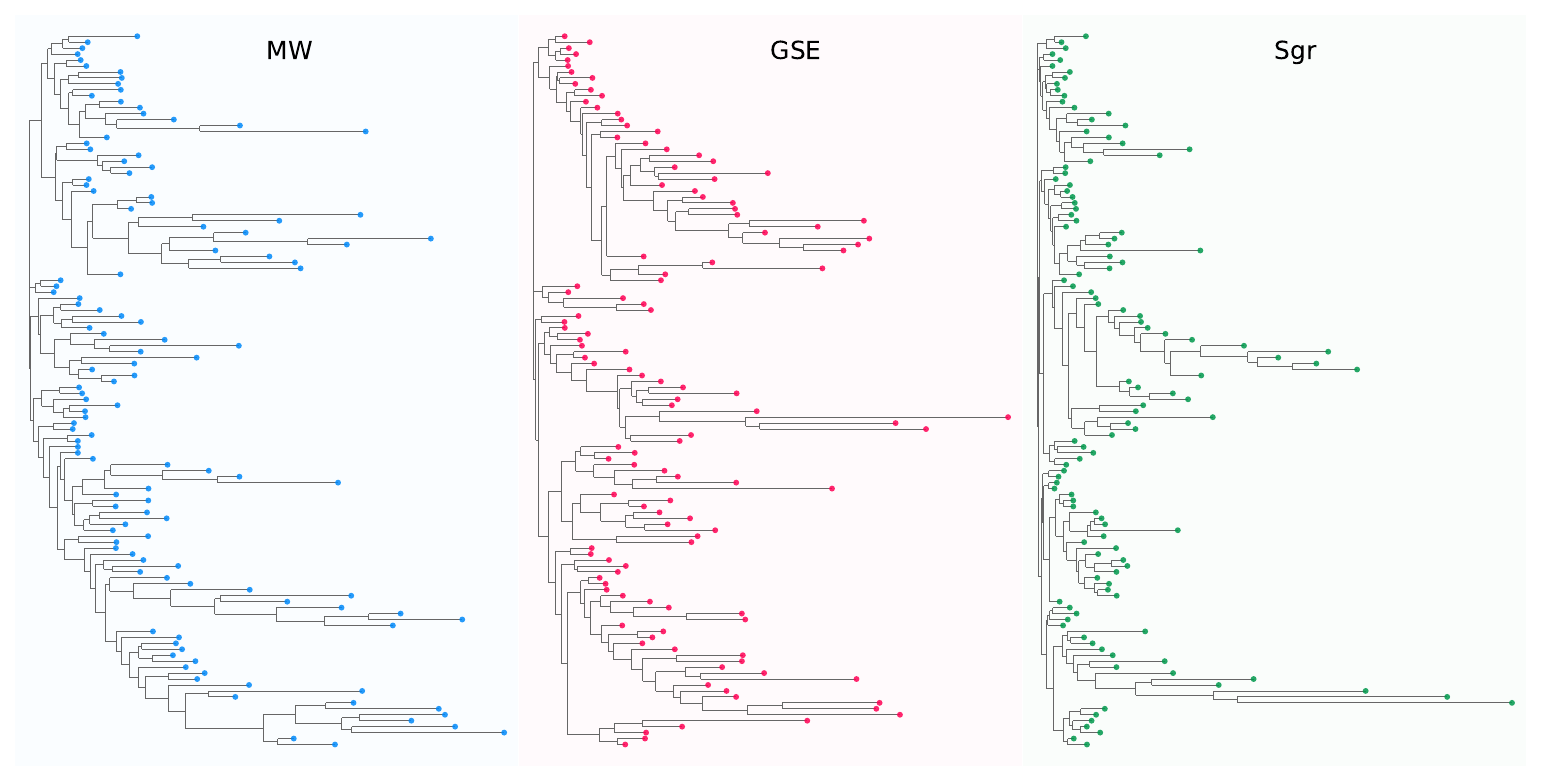}
\caption{Representative phylogenetic trees built from random subsamples of the Milky Way (left), Gaia--Sausage--Enceladus (center), and Sagittarius (right) populations. Branch lengths encode chemical distance between nodes. The three trees are built from disjoint sets of stars and cannot be compared by classical leaf-matching metrics; the graph-based measures developed in this work quantify their structural differences without requiring shared leaves.}
\label{fig:trees}
\end{figure*}

Figure~\ref{fig:trees} shows examples of phylogenetic trees constructed from random subsamples of the MW, GSE, and Sgr populations. As described in Sect.~\ref{sec:methods}, each leaf corresponds to an individual star and branch lengths encode chemical divergence between nodes. 
Some qualitative features are visible by eye. The MW tree separates into two broad groupings, each internally organized into smaller clusters; the GSE tree instead shows several comparably sized clusters distributed along its length, including one notably long branch; the Sgr tree shows shorter, densely packed internal branches, with branch lengths increasing progressively toward the leaves. Like standard clustering algorithms, which return a partition regardless of whether the data possess genuine cluster structure, Neighbor-Joining always returns a fully resolved binary tree: the branches and apparent clusters visible here are not by themselves evidence of physically distinct sub-populations. Associating such features (the number and depth of clusters, the distribution of branch lengths) with the coexistence of chemically distinct sub-populations with different evolutionary histories would therefore require additional information not used here, such as stellar ages or kinematics. We leave this interpretation to future work and focus here on establishing a quantitative framework for comparing tree structure.

Each tree is constructed from a distinct set of stars, so the three share no common leaves and cannot be placed in correspondence by the classical leaf-matching metrics of Sect.~\ref{sec:tree_metrics}. They also illustrate the practical difficulty of comparing stellar phylogenies quantitatively: with roughly one hundred leaves per tree, visual inspection can reveal only coarse structural impressions, not a measurement of how much two trees differ, nor a way to rank or compare trees built from different stars. Both observations motivate label-invariant spectral measures.
\subsection{Solar twins from \citet{walsen2024assembling}}
This dataset consists of solar twins from \citet{walsen2024assembling}, with chemical abundances derived from GALAH DR3 spectra. We use the same sample of 201 stars analyzed in their work, restricted to high-quality abundance measurements with typical uncertainties of $\approx 0.026$ dex.

\subsection{Milky Way, Gaia--Sausage--Enceladus, and Sagittarius populations}
We consider stellar populations associated with the Milky Way (MW), Gaia–Sausage–Enceladus (GSE), and the Sagittarius dwarf spheroidal galaxy (Sgr dSph).
The Sgr sample was selected to cover the core of the galaxy while excluding the metal-poor nuclear stellar cluster M54. Candidate members were taken from \citet{2022Vitali} and cross-matched with APOGEE DR17 \citep{2022ApJS..259...35A}.

MW and GSE members were selected from the chemo-kinematical catalog of \citet{2024Fiorentin}. To ensure homogeneous abundance measurements across all populations, we retained only stars with reliable abundance estimates (\texttt{X\_FE\_FLAG = 0} for all abundances used in the analysis) and no significant data reduction issues (\texttt{STARFLAG = 0}).
This selection resulted in a total of $354$ stars for Sgr, $288$ for GSE, and $\approx 160\, 000$ for the MW, with median abundance uncertainties in the range $0.02$--$0.03$ dex.

\subsection{Solar twins from \citet{nissen2020high}}
This dataset consists of 70 solar-type stars from \citet{nissen2020high}, located in the solar neighbourhood within $60$ pc of the Sun, with very high-precision abundances (SNR > 600) and average statistical uncertainties of order $\approx 0.01$ dex (their Table 3).
The age-metallicity distribution of the sample reveals two chemically distinct stellar populations (41 older and 29 
younger stars), interpreted by \citet{nissen2020high} as evidence for two major episodes of gas accretion and star formation in the Galactic disk.

\section{Results}\label{sec:results}
We now present the results of the benchmarks introduced in Sect.~\ref{sec:benchmark}, evaluating how different distance measures perform in terms of (1) sensitivity to perturbations in chemical data and (2) ability to discriminate between distinct evolutionary histories.
We adopt $\tau = 0.2$ dex throughout (the scale parameter of the affinity kernel, Eq.~\ref{eq:kernel}), chosen to avoid compressing all edge weights toward unity; the choice is discussed and validated in Appendix~\ref{app:compression}.
\subsection{Consistency within the same evolutionary history}
\label{sec:results-robustness}
\begin{figure}
    \centering
    \includegraphics[width=\columnwidth]{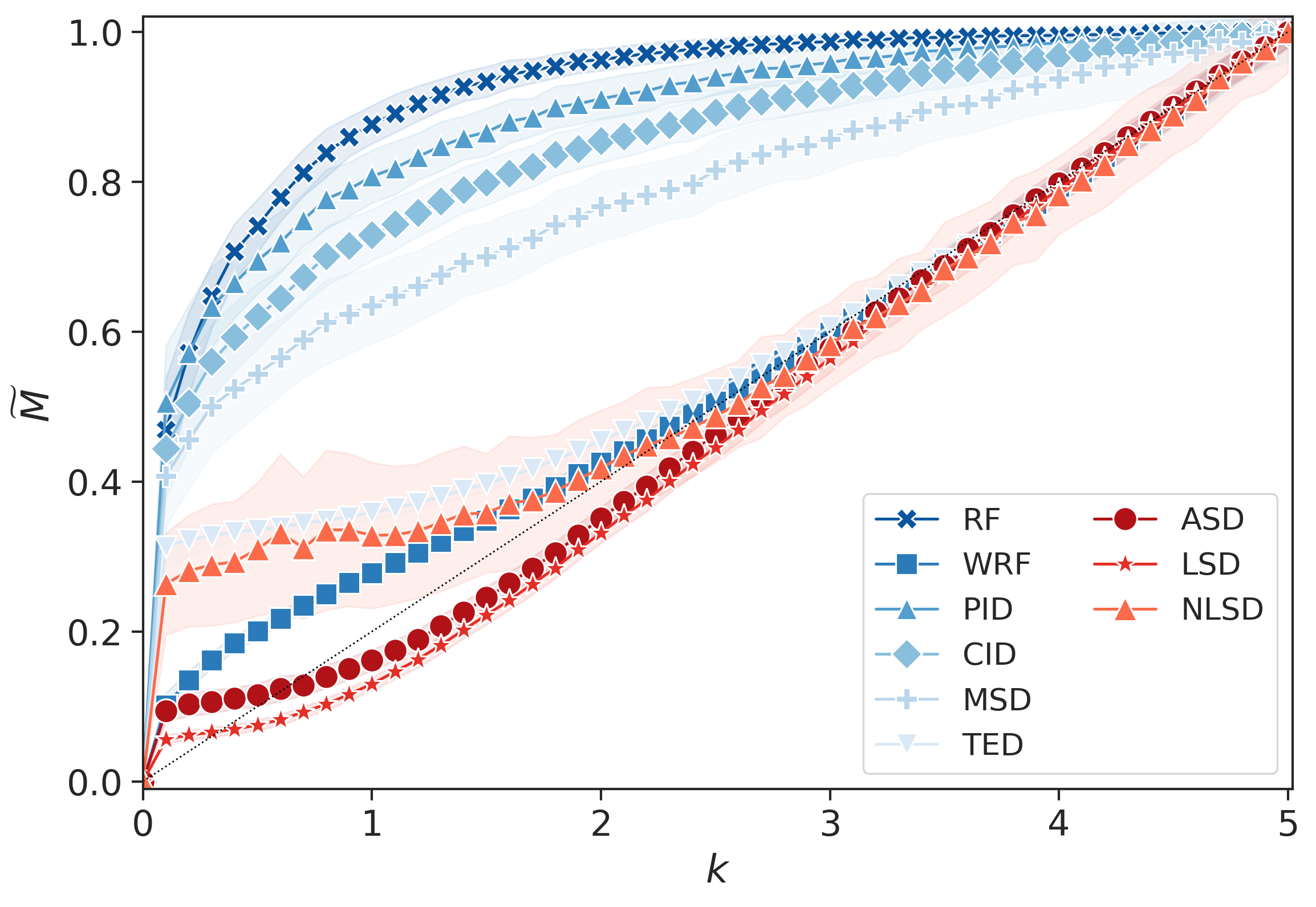}
    \caption{Rescaled measure values $\widetilde{M}$ as a function of noise level, measured relative to a fixed reference tree constructed from 201 solar twins from \cite{walsen2024assembling}. Each point represents the average over 200 realizations for a given $k \in [0,5]$. Tree-based metrics are shown in shades of blue, graph-based measures in red. All values are normalized to their maximum value. The dotted black line shows the reference slope $\widetilde{M} = k/5$.}
    \label{fig:perturbation}
\end{figure}

To assess how each distance measure responds to small variations in input data, we analyzed the distribution of distances between reference trees and trees reconstructed from perturbed versions (Eq.~\ref{eq:noisy}, Sect.~\ref{subsec:sensitivity}) of the same chemical abundance matrix of 201 solar twins (\citealt{walsen2024assembling}). Figure~\ref{fig:perturbation} shows the mean (solid line) and standard deviation (shaded region) of these distances as a function of the noise parameter~$k$ (Eq.~\ref{eq:noisy}), with tree-based metrics shown in blue and graph-based measures in red.
As $k$ increases, the perturbations become large enough to obscure the evolutionary signal, and the resulting trees begin to diverge more significantly from the reference.

Tree-based metrics such as RF, PID, CID, and MSD show a steep increase at small $k$, followed by early saturation, often reaching their maximum values around $k \lesssim 2$ (corresponding to perturbations of order $2\sigma$, or $\approx0.05$ dex for a typical element in this sample); these metrics lose discriminative power, as distances rapidly approach their maximum values.
In contrast, spectral distances such as ASD and LSD, as well as the WRF metric, show a more gradual increase, with sublinear growth at small $k$ and an approximately linear trend at larger perturbations.
The absence of saturation in graph-based measures suggests that they maintain sensitivity across a wide range of perturbation levels.

Also, as shown in Fig. \ref{fig:perturbation}, the NLSD increases more sharply than the other spectral measures at low noise levels ($k<1$). 
Since the NLSD rescales edge weights by node degree, it suppresses the influence of overall edge-weight scale relative to LSD, making it more sensitive to changes in the relative organization of the weighted graph than to changes in edge-weight magnitude. This suggests that NLSD is more sensitive to noise-induced structural rearrangements. LSD and ASD, by comparison, retain greater sensitivity to edge-weight differences.

\subsection{Discriminative power across evolutionary histories}
\label{sec:results-histdiff}
\begin{figure*}
\centering
\includegraphics[width=\textwidth]{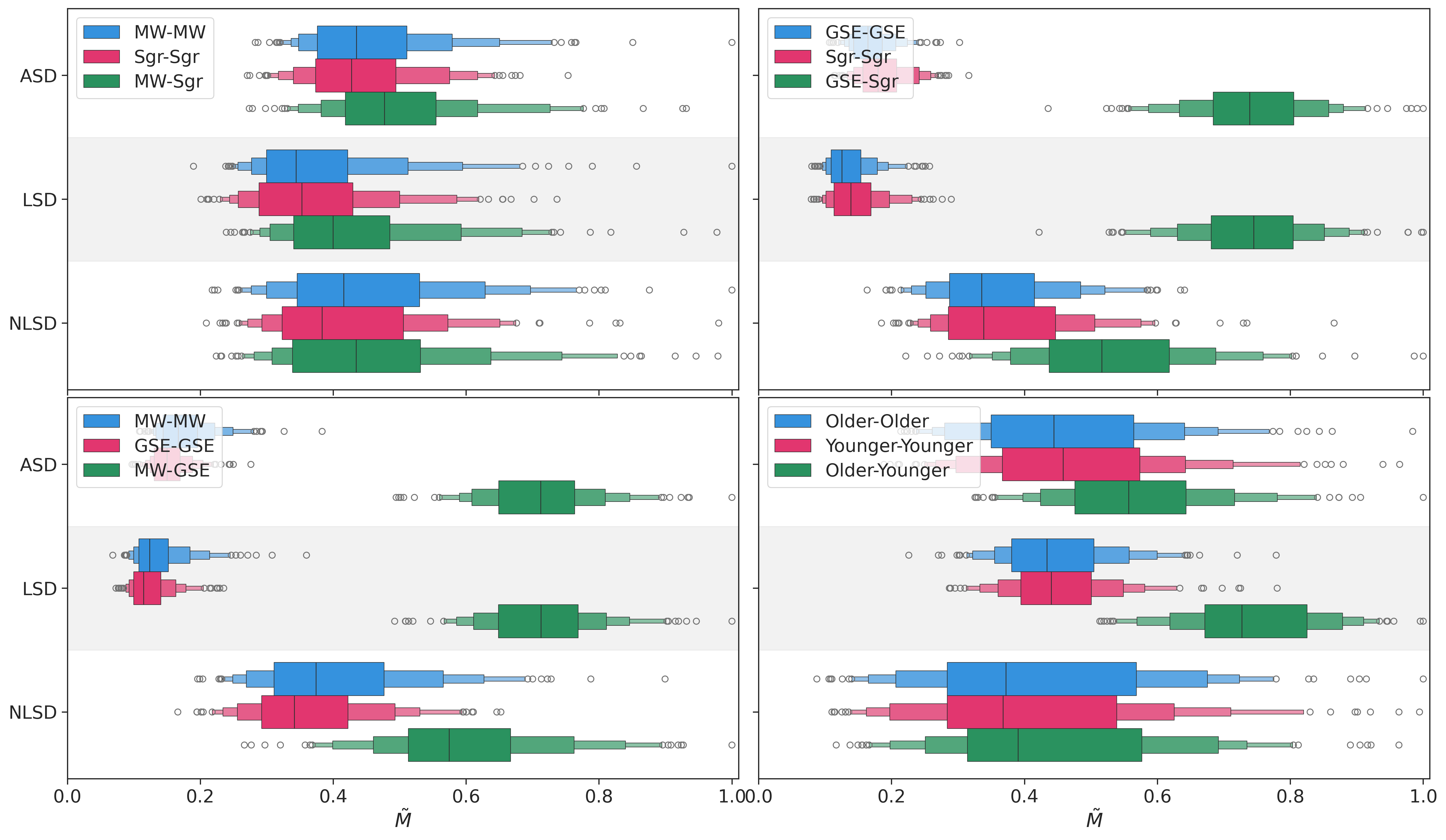}
\caption{Distributions of rescaled graph-based distance measures $\widetilde{M}$ between phylogenetic trees constructed from different stellar populations (Sect.~\ref{sec:data}). Each panel corresponds to a set of populations (e.g., MW, GSE, Sgr), treated as distinct evolutionary histories. Distances within the same population (intra-history) are shown in blue and red, while distances between populations (inter-history) are shown in green. Results are reported for ASD, LSD, and NLSD. The relative separation between intra- and inter-history distributions reflects the discriminative ability of each measure.}
\label{fig:historiesboxen}
\end{figure*}
\begin{table}
\caption{AUC values for the separation between intra- and 
inter-population distance distributions.}
\label{tab:auc}
\centering
\begin{tabular}{lcccc}
\hline\hline
Measure & GSE--Sgr & MW--GSE & MW--Sgr & Nissen \\
\hline
ASD & 1.00 & 1.00 & 0.62 & 0.70 \\
LSD & 1.00 & 1.00 & 0.64 & 0.98 \\
NLSD & 0.84 & 0.88 & 0.55 & 0.54 \\
\hline
\end{tabular}
\tablefoot{Values near 1 indicate clean separation between intra- and inter-population distributions; values near 0.5 indicate no discriminative power.}
\end{table}
We evaluate the ability of graph-based distances to distinguish trees derived from different stellar populations, which we interpret as realizations of distinct evolutionary histories (Sect.~\ref{sec:data}), under the scope established in Sect.~\ref{sec:benchmark}.
Since classical tree metrics require identical leaf sets, we restrict the analysis to spectral measures.

For each pair of populations, we compare inter-population distances to intra-population variability. The two are estimated differently depending on the dataset.
For the \citet{nissen2020high} sample, where the number of stars is limited, we estimate intra-population variability by generating perturbed realizations of the abundance matrix using Gaussian noise. We adopt the average statistical ($1\sigma$) errors reported in their Table~3, which are of order $\sim 0.01$ dex. The two populations differ in size (41 older and 29 younger stars); to avoid size-induced structural differences between trees, we construct all trees from random subsamples of 29 stars, matching the size of the smaller population.

For the MW, GSE, and Sgr populations, which have larger and very different pool sizes ($\approx 160\, 000$, $288$, and $354$ stars respectively) and metallicity distributions, intra-population variability is estimated from random subsamples of the data. For MW–Sgr, we estimate the target Sgr [Fe/H] distribution with a kernel density estimate and resample the MW sample to that density; for MW–GSE, we apply an upper [Fe/H] cut to the MW sample at the maximum metallicity of the GSE sample. Both operations primarily match the metallicity coverage of the two populations, and as a side effect also reduce the effective MW pool to a size closer to that of Sgr and GSE, limiting the extent to which subsampling variability is artificially suppressed by MW's much larger pool. For GSE–Sgr, the two pools are already of comparable size, so no matching is applied. In all cases, trees are then constructed from random subsamples of 120 stars.
For the MW, GSE, and Sgr populations no noise is added; the intra-population distributions discussed here therefore isolate the effect of subsampling different stars from the same population, independent of measurement uncertainty.

Results are shown in Fig.~\ref{fig:historiesboxen}, with blue and red denoting intra-population variability and green inter-population variability. The corresponding AUC values quantify this separation as the probability that a randomly drawn inter-population distance exceeds a randomly drawn intra-population distance \citep{hanley1982meaning}. The resulting values are given in Table~\ref{tab:auc}; values near one indicate clean separation, whereas values near 0.5 indicate little or no discriminative power.

ASD and LSD achieve perfect separation for the chemically distinct MW--GSE and GSE--Sgr pairs (AUC=1 for both), but their performance decreases for the chemically similar MW--Sgr pair (AUC=0.62 and 0.64 respectively). LSD also outperforms ASD for the Nissen older--younger pair (AUC=0.98 vs 0.70). This result suggests that global structural organization is more informative than fine-scale branch-length variation when populations are chemically similar. Overall, LSD provides the most consistent discrimination across the population pairs considered here.

The normalized Laplacian is invariant to a uniform rescaling of edge weights. It therefore suppresses sensitivity to the overall scale of chemical divergence but retains sensitivity to its relative organization. Its near-chance performance for MW--Sgr (AUC=0.55) indicates that, after suppressing the overall scale of the weighted trees, the remaining structural differences are too small to distinguish from subsampling variability. Conversely, NLSD retains substantial separation for GSE--Sgr and MW--GSE (AUC=0.84 and 0.88), indicating that the organization of the corresponding weighted trees remains distinguishable after this normalization.

For the MW, GSE, and Sgr populations specifically, since no noise was added, these AUC values directly reflect the impact of subsampling: variability from drawing different stars is small compared to the MW–GSE and GSE–Sgr differences, but comparable to the MW–Sgr difference.

A consistent ordering is observed for ASD and LSD: distances between Milky Way and GSE populations are the largest, followed by GSE–Sgr pairs, while MW–Sgr pairs show smaller separation. 
This ordering places GSE furthest from the MW and Sgr closer to the MW in the spectral-distance space. The ordering is consistent with current astrophysical expectations \citep{hasselquist2021apogee,vitali2025pristine}.
GSE has the highest early star formation efficiency among MW satellites, with star formation abruptly truncated as a consequence of its merger with the Milky Way \citep{helmi201merger, 
belokurov2018coformation}, resulting in a distinct $[\alpha/\text{Fe}]$--$[\text{Fe/H}]$ sequence that separates from the Milky Way at $[\text{Fe/H}] \approx -1.2$, alongside clear deficiencies in $[\text{Al/Fe}]$ and $[\text{Ni/Fe}]$.

Sgr, whose progenitor is estimated to be quite massive \citep[$\approx 10^{10-11}M_\odot$,][]{lokas2024sagittarius}, experienced a more prolonged star formation history than GSE, showing intermediate star formation efficiency and metal-poor stars whose chemical abundances overlap closely with those of the Milky Way halo \citep{vitali2025pristine}.
This greater overlap with Milky Way patterns is consistent with the smaller MW–Sgr separations recovered by our spectral measures.

\section{Conclusions and discussion}\label{sec:conclusions}
We tested whether phylogenetic trees built from stellar chemical abundances can be compared in a way that distinguishes populations with different evolutionary histories, while remaining robust to noise and sampling variation.
Classical tree comparison metrics are limited in this setting: they require identical leaf sets, and their behavior under observational noise and varying samples is not well characterized. This motivates the need for more flexible and informative measures.
We reframed tree comparison as a graph comparison problem. Phylogenetic trees are fundamentally graphs, and treating them explicitly as such allows us to leverage spectral properties to compare their structural organization, even when trees are built from different sets of stars.

In the first benchmark, we assessed the sensitivity of each measure to perturbations in the input data, mimicking observational uncertainty within the same evolutionary history. Tree-based metrics saturated quickly and lost discriminative power beyond small noise levels ($\approx2\sigma$, or $\approx0.05$ dex for a typical element). In contrast, spectral graph distances showed smooth, monotonic responses to increasing perturbations, maintaining sensitivity across the full range of noise levels without early saturation.

In the second benchmark, we evaluated whether spectral distances distinguish trees derived from stellar populations used as proxies for different evolutionary histories. LSD provides strong separation for MW--GSE, GSE--Sgr, and the Nissen older--younger comparison, whereas separation is more modest for MW--Sgr, where the intra- and inter-population distributions overlap substantially. ASD performs well for the chemically distant MW--GSE and GSE--Sgr pairs, but shows reduced separation for both MW--Sgr and the Nissen older--younger comparison.
The recovered ordering, with MW--GSE separations being the largest, followed by GSE--Sgr, and MW--Sgr the smallest, is consistent with the known chemical enrichment histories of these populations.
NLSD, invariant to a uniform rescaling of edge weights, retains substantial separation for GSE--Sgr and MW--GSE. This indicates that differences in the relative organization of the weighted trees remain detectable for GSE even after the overall chemical scale is suppressed.

The two benchmarks probe complementary sources of variability within a shared evolutionary history: perturbations in input data and subsampling of stellar populations.
Spectral distances remain informative under both sources of variability and distinguish the pre-defined populations considered here. These results suggest that spectral graph measures provide a promising framework for comparing phylogenetic trees in astrophysical contexts, where uncertainty and incomplete sampling are common.

While tree-to-tree comparisons offer a useful tool for studying chemical evolution, their effectiveness ultimately depends on the quality of the inferred trees. Improved tree-building methods, whether via Bayesian inference, machine learning, or domain-specific heuristics, could yield more faithful graph structures and, by extension, more reliable comparisons. We also note that the framework implicitly assumes each input dimension traces genuine chemical enrichment rather than an observational or instrumental confound. A noisy or weakly informative abundance, or one affected by residual correlations with stellar parameters or analysis systematics, can alter chemical distances and inferred tree structure. Careful abundance selection, quality control, and treatment of measurement uncertainties are therefore prerequisites for applying this framework to new datasets.

Nevertheless, our results indicate that trees built with a standard method such as Neighbor-Joining already contain sufficient structural information for meaningful spectral comparison.
These results are not intended to show that spectral tree comparison outperforms direct comparison of the underlying chemical-abundance distributions: population membership is, after all, already evident from the abundance distributions themselves (Fig.~\ref{fig:afe_all}). Instead, they demonstrate that a label-invariant representation of tree structure retains discriminative information when the trees contain different stars. The spectral measures also respond differently to global and fine-scale structural variation. The fundamental question, which we leave for future work, is whether the branching structure of these trees encodes information about the underlying enrichment process that is not readily apparent from abundance-plane representations alone. A label-invariant method for comparing tree structure provides a methodological basis for addressing this question.

Understanding how this depends on the choice of tree-construction algorithm and chemical distance metric remains an important direction for future work, particularly before broader astrophysical conclusions can be drawn.
This positions spectral tree comparison as a structural summary that complements rather than substitutes direct abundance comparison, useful where a representation independent of stellar ages and explicit enrichment-model fitting is desired.
More broadly, this work is one step in an ongoing effort to develop the tools needed for phylogenetic analyses of the Galaxy, moving from direct abundance-space comparisons and chemical tagging toward the use of hierarchical structure to study chemical evolution.

Finally, scalability may become a practical consideration for larger samples. In the present implementation, standard Neighbor-Joining tree construction and the full eigendecomposition required by the spectral distances both have cubic scaling with sample size. This is manageable for the \(N \approx 100\) stars per tree considered here, but would become costly at catalog scale. Future large-sample applications could use approximate tree construction, representative subsampling, or truncated and randomized eigensolvers, accepting that partial spectra may discard some fine-scale structural information.

This framework opens several directions for further work. 
One is to make spectral comparisons more interpretable, complementing their sensitivity and leaf-invariance. Spectral distances quantify overall structural differences between trees but do not identify which structural properties drive them. This can be addressed in two ways. First, by pairing spectral distances with independent, targeted tree summaries: scalar descriptors such as the Corrected Colless index isolate a single interpretable property, tree imbalance, which \citet{tapia2026reconstructing} associate with the burstiness of the underlying star formation history in simulated disc galaxies. Second, by looking inside the spectrum itself: individual eigenvectors, rather than the full eigenvalue distance, can carry structural meaning; for instance, the Fiedler vector (the eigenvector associated with the second-smallest Laplacian eigenvalue) encodes large-scale partitioning of the graph, and could offer a more detailed characterization of evolutionary structure than the aggregate distance alone. Combining global spectral comparisons with such targeted or component-level statistics could connect the discriminative power demonstrated here to specific astrophysical processes.

Spectral measures and properties could be used to benchmark tree-construction algorithms, identify informative chemical dimensions, or explore enrichment pathway diversity in simulations. A natural application is to compare an observed tree against trees built from simulated chemical enrichment histories, to identify which simulated scenario best reproduces the observed structure.
Graph-based descriptors could also serve as inputs to machine learning models. For example, supervised or contrastive approaches can learn to separate trees from distinct histories when population labels are available.
These directions point toward task-specific models grounded in phylogenetic structure for studying the chemical evolution of the Milky Way.

\section*{Data availability}
All code accompanying this manuscript is available on GitHub at \url{https://github.com/theosig/phylographs}.
\begin{acknowledgements}
   TS acknowledges financial support from Inria Chile ANID project CTI230007. PJ acknowledges partial financial support of FONDECYT Regular Grant Number 1231057. C.A.G. acknowledges support from Agencia Nacional de Investigación y Desarrollo (ANID) through FONDECYT Regular 1262342. PBT acknowledges fondecyt Regular 1240465 and ANID Basal Project FB210003. B.T.C. gratefully acknowledges funding by ANID (Beca Doctorado Nacional, Folio 21232155). The authors thank Luís Llanca, Álvaro Márquez, and Hernan Lira for their contributions during the initial stages of this project, and the referee for their constructive comments on the paper.
\end{acknowledgements}
\bibliographystyle{aa}
\bibliography{references}
\begin{appendix}
   \section{Tree-to-tree distance measures}\label{app:tree_metrics}
   
   This appendix provides formal definitions of the classical tree-to-tree distance measures discussed in Sect.~\ref{sec:tree_metrics}. These are standard dissimilarity measures used to compare phylogenetic trees.
   
   \paragraph{Robinson--Foulds distance.}
   The Robinson--Foulds (RF) distance \citep{RobinsonFoulds1981} is a topological measure that quantifies the dissimilarity between two trees by counting unmatched bipartitions (splits).  
   Given two unrooted trees \(T_1\) and \(T_2\) defined on the same set of leaves, the RF distance is:
   \[
   M_{\mathrm{RF}}(T_1, T_2) = |C(T_1) \setminus C(T_2)| + |C(T_2) \setminus C(T_1)|,
   \]
   where \(C(T)\) denotes the set of non-trivial splits (splits that partition more than one leaf on each side) induced by the internal edges of tree \(T\).  
   For unrooted binary trees with \(n\) leaves, the RF distance ranges from 0 (identical topologies) to \(2(n - 3)\).  
   This measure depends on leaf labels and ignores branch lengths.
   
   \paragraph{Weighted Robinson--Foulds distance.}
   The weighted Robinson--Foulds (WRF) distance \citep{RobinsonFoulds1981} extends RF by incorporating branch lengths:
   \[
   M_{\mathrm{WRF}}(T_1, T_2) = \sum_{s \in C(T_1) \cup C(T_2)} \left| l_1(s) - l_2(s) \right|,
   \]
   where \(l_i(s)\) denotes the length associated with split \(s\) in tree \(T_i\), and \(l_i(s)=0\) if the split is absent from \(T_i\).  
   This formulation accounts for both topological differences and discrepancies in branch lengths.
   
   \paragraph{Matching Split distance.}
   The matching split distance (MSD; \citealt{Bogdanowicz2012}) measures dissimilarity by optimally pairing splits from two trees. It is defined as:
   \[
   M_{\mathrm{MS}}(T_1, T_2) = \min_{M} \sum_{(s,t)\in M} h_S(s,t),
   \]
   where \(M\) is a matching between splits of \(T_1\) and \(T_2\) that minimizes the total cost, and the dissimilarity between two splits \(A_1 \mid B_1\) and \(A_2 \mid B_2\) is:
   \[
   h_S(A_1 \mid B_1, A_2 \mid B_2) = \min\{\, |A_1 \Delta A_2|,\; |A_1 \Delta B_2| \,\},
   \]
   with \(\Delta\) denoting the symmetric difference.  
   This measure captures partial similarity between splits rather than exact agreement.
   
   \paragraph{Clustering information distance.}
   The clustering information distance (CID) \citep{Llabres2021} compares trees using information-theoretic quantities defined on their splits:
   \[
   M_{\mathrm{CID}}(T_1, T_2) = \frac{1}{2} \left( \sum_{S \in C(T_1)} H(S) + \sum_{S \in C(T_2)} H(S) \right) - \mathrm{MCI}(T_1,T_2),
   \]
   where \(H(S)\) is the entropy associated with the bipartition induced by split \(S\), and \(\mathrm{MCI}(T_1,T_2)\) is the Mutual Clustering Information:
   \[
   \mathrm{MCI}(T_1,T_2) = \sum_{(S_1,S_2)\in M} I_{\mathrm{Cl}}(S_1; S_2).
   \]
   Here \(M\) denotes the matching between splits that maximizes the shared information, and
   \[
   I_{\mathrm{Cl}}(S_1; S_2) =
   \sum_{(X,Y)\in \{A_1,B_1\}\times\{A_2,B_2\}}
   P_{\mathrm{Cl}}(X,Y)\log\frac{P_{\mathrm{Cl}}(X,Y)}{P_{\mathrm{Cl}}(X)P_{\mathrm{Cl}}(Y)},
   \]
   where \(X\) is the full set of leaves, \(P_{\mathrm{Cl}}(A)=|A|/|X|\) and \(P_{\mathrm{Cl}}(A_1,A_2)=|A_1 \cap A_2|/|X|\).
   This measure captures both exact and partial agreement between tree partitions.
   
   \paragraph{Phylogenetic information distance.}
   The phylogenetic information distance (PID) \citep{Llabres2021} is an information-theoretic measure defined as:
   \[
   M_{\mathrm{PID}}(T_1,T_2) = H(T_1) + H(T_2) - 2\, \mathrm{MPI}(T_1,T_2),
   \]
   where \(H(T)\) denotes the total phylogenetic information content of tree \(T\), and \(\mathrm{MPI}(T_1,T_2)\) is the mutual phylogenetic information shared between the two trees.  
   This measure quantifies the amount of shared evolutionary information between trees.
   
   \paragraph{Tree edit distance.}
   The tree edit distance (TED) measures the minimum cost of transforming one tree into another through a sequence of edit operations: node insertion, deletion, and relabeling. Given two trees $T_1$ and $T_2$, it is defined as:
   \begin{equation}
       M_{\mathrm{TED}}(T_1, T_2) = \min_{\mathcal{E}} 
       \sum_{e \in \mathcal{E}} c(e),
   \end{equation}
   where $\mathcal{E}$ is the set of edit operations transforming $T_1$ into $T_2$ under an optimal alignment, and $c(e) \geq 0$ is the cost assigned to each operation. In this work, edit costs are defined as a function of branch lengths: insertion and 
   deletion of a node $v$ each carry cost $w_v$ (the branch length of that node), while relabeling node $v$ to node $u$ carries 
   cost $|w_v - w_u|$. Under this formulation, TED is sensitive to metric structure and penalizes both topological differences and 
   branch-length discrepancies. The distance is computed using the algorithm of \citet{zhang1989simple}.

   \section{Sensitivity to graph construction choices}
   \label{app:compression}
   \begin{figure}[h]
       \centering
       \includegraphics[width=\columnwidth]{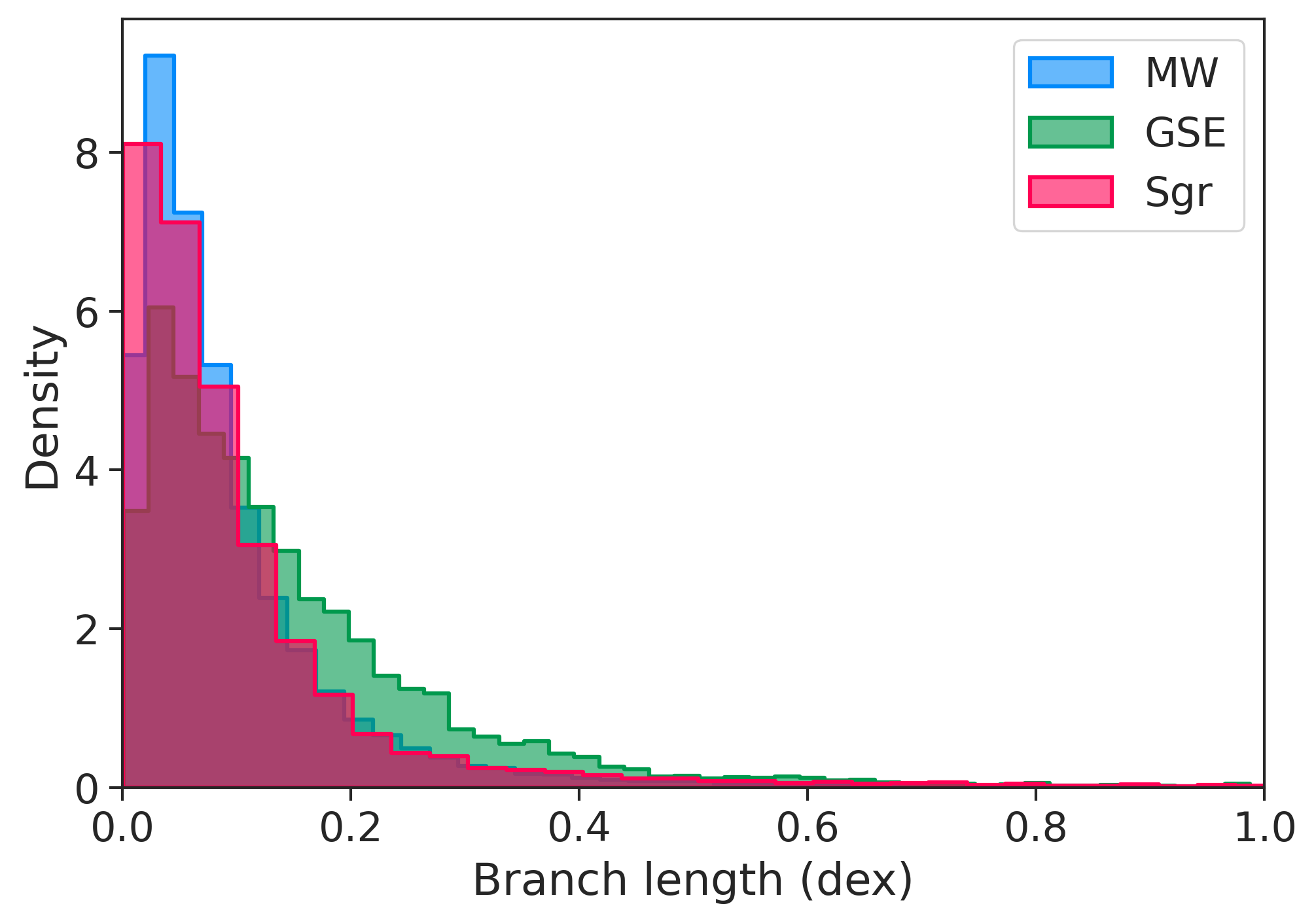}
       \caption{Distribution of branch lengths in trees built from the
       Milky Way, GSE, and Sagittarius samples ($L_1$ chemical distance).}
       \label{fig:branchlengths}
   \end{figure}
   
   \begin{figure*}
       \centering
       \includegraphics[width=\textwidth]{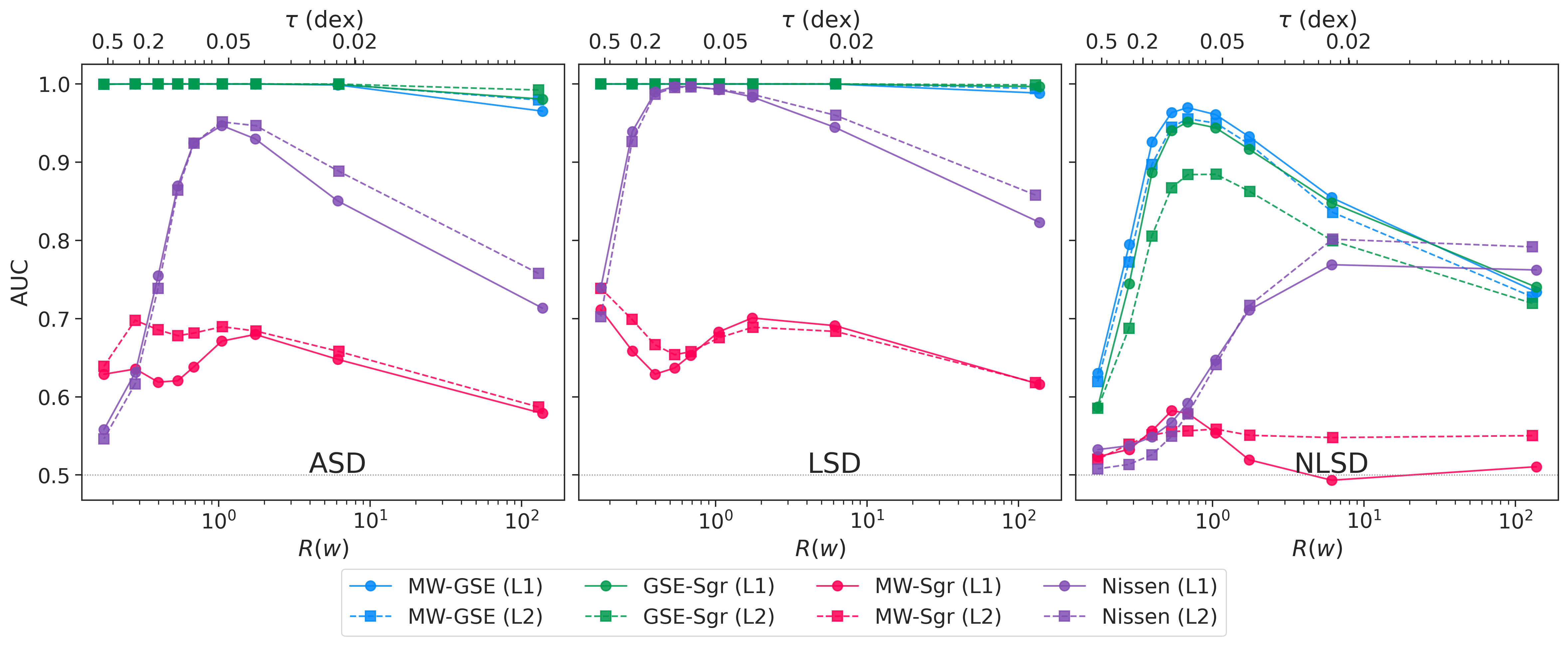}
       \caption{AUC between intra- and inter-population distance distributions as a function of the relative weight spread $R(w)$ (Eq.~\ref{eq:spread}), for each spectral measure (panels) and population pair (colors). Solid lines with circles use the $L_1$ chemical distance, dashed lines with squares the $L_2$ distance. The upper axis shows the corresponding $\tau$, calibrated to the $L_1$ branch-length distribution. The dotted line marks AUC $=0.5$ (no separation).}
       \label{fig:auc-compression}
   \end{figure*}
   
   The graph representation of Sect.~\ref{sec:methods} depends on several construction choices; here we examine two of particular relevance for this work: the decay parameter $\tau$ of the affinity kernel (Eq.~\ref{eq:kernel}) and the metric used to compute chemical distances between stars, here the Manhattan ($L_1$) distance. We show that these are not independent: both act on the spectral comparison through a single effective quantity, the degree to which the edge weights are spread across their available range.
   
   The affinity weights $w_{ij} = e^{-d_{ij}/\tau}$ map branch lengths $d_{ij} \in [0,\infty)$ onto $(0,1]$, with $\tau$ setting the scale over which affinity decays. Their effect depends on the branch-length distribution, which for our trees is peaked at small values (median $\mathcal{O}(10^{-1})$ dex with a tail extending to $\sim 1$ dex, Fig.~\ref{fig:branchlengths}). 
   When $\tau$ is much larger than the typical branch length, $e^{-d_{ij}/\tau} \approx 1$ for nearly all edges: the kernel is flat and the weighted graph approaches an unweighted one, suppressing branch-length information. When $\tau$ is much smaller than the typical branch length, the longest branches are driven toward zero weight, partially disconnecting the graph and producing near-degenerate eigenvalues that reflect numerical disconnection rather than tree structure. We summarise this with the relative spread of the weights,
   \begin{equation}
       R(w) = \frac{\mathrm{std}(w_{ij})}{\mathrm{median}(w_{ij})},
       \label{eq:spread}
   \end{equation}
   which is large when most weights are compressed near zero (small $\tau$) and small when they are compressed near unity (large $\tau$).
   
   Figure~\ref{fig:auc-compression} shows the AUC for each spectral measure and population pair as a function of $R(w)$, recomputed over a range of $\tau$. The discriminative power is non-monotonic in $\tau$ (equivalently, in $R(w)$): it is low both when the weights are nearly uniform (large $\tau$, small $R$) and when they are dominated by a small subset of differentiated edges (small $\tau$, large $R$), and peaks at intermediate spread wherever a measure is not already saturated. For the chemically distinct MW--GSE and GSE--Sgr pairs, ASD and LSD remain close to AUC$=1$ across nearly the full range of $R(w)$, while NLSD displays a clear rise and fall with a maximum at intermediate spread. For the harder MW--Sgr and Nissen pairs, where no measure saturates, ASD and LSD show the same intermediate peak instead, while NLSD remains comparatively flat and close to chance throughout. The AUC-optimal scale lies near $\tau \approx 0.1$ dex; at this value, however, the longest branches in our trees map to weights as small as $w \sim 10^{-12}$, leaving little margin before individual edges become numerically negligible. We thus adopt the slightly larger $\tau = 0.2$ dex as a conservative regularization choice, which still lies on the discriminability plateau in Fig.~\ref{fig:auc-compression} while keeping all edge weights at a comfortably resolvable scale (minimum weight $\sim 10^{-6}$ across all trees considered), and is adopted throughout the main text.
   
   The same figure also isolates the role of the chemical distance metric. The $L_2$ (Euclidean) distance changes the scale of the branch-length distribution, and hence the weight spread at a given $\tau$. To separate this from a direct effect of the metric, we repeat the analysis for both $L_1$ and $L_2$. At matched spread, $L_1$ and $L_2$ trace a common curve for ASD and LSD for every population pair, with a mean $|\mathrm{AUC}_{L_1}-\mathrm{AUC}_{L_2}|$ of $0.001$--$0.002$ for MW--GSE and GSE--Sgr and below $0.02$ for MW--Sgr and Nissen under either measure. The discrepancy is largest for the chemically similar MW--Sgr and Nissen pairs, where the underlying separation signal is weaker and the comparison is correspondingly more sensitive to second-order effects of the metric. NLSD shows somewhat larger discrepancies (up to $0.05$, for GSE--Sgr), reflecting its weaker and less stable overall performance. The metric choice therefore carries little independent discriminative information: its influence is mediated almost entirely by the weight spread it induces, once $\tau$ is calibrated to the branch-length scale of the data.

   We also assess sensitivity to the choice and scaling of the input chemical dimensions. Abundances are used as raw [X/Fe] values without additional standardization (Sect.~\ref{sec:trees}); we test this choice in two ways, using the four population pairs of Sect.~\ref{sec:results}.
   First, we standardize each element (zero mean, unit variance, computed jointly over the two populations in each pair) before tree construction. Standardization changes the branch-length scale substantially (median $\approx 0.71$ vs.\ $\approx 0.08$\,dex for the raw case), so reusing $\tau=0.2$\,dex would drive edge weights to numerically negligible values; we instead recalibrate the kernel scale to $\tau_{\rm std}=1.5$, close to the value that reproduces the raw-data $R(w)$ plateau, and use this value throughout the standardized runs (Table~\ref{tab:sensitivity-standardization}).
   
   Second, we repeat the full analysis with each of the seven elements removed in turn, using $\tau=0.2$\,dex and the same seven-element complete-case star pools throughout (Table~\ref{tab:sensitivity-loo}).
   Both tests show larger movement for the two harder pairs (MW--Sgr, Nissen) than for the already-saturated MW--GSE and GSE--Sgr, where ASD and LSD remain at or near $\mathrm{AUC}=1$ throughout. Among the three measures, LSD is the most stable overall, with a maximum AUC shift of $0.20$ across both tests (leave-one-out, MW--Sgr); ASD is comparable, with a maximum shift of $0.30$ (standardization, Nissen).
   NLSD is the most volatile in both tests, with shifts of up to $0.39$ under standardization and $0.22$ under single-element removal, consistent with its greater sensitivity to changes in the relative organization of the weighted graph and to topological rearrangements, as noted in Sects.~\ref{sec:results-robustness} and \ref{sec:results-histdiff}. 
   Overall, these tests show that the discriminative power of ASD and LSD is not driven by the raw-abundance representation or by any single chemical dimension.
   
   This aggregate robustness does not imply that the elements are interchangeable: different elements trace different aspects of chemical evolution and can affect the inferred tree structure in different ways. The pattern of AUC changes under single-element removal may therefore itself contain astrophysical information about which abundances are most diagnostic of a given evolutionary history, for instance by identifying which chemical dimensions best separate specific populations. Systematically exploring this is a natural extension of the present framework and is left for future work.
   
   \begin{table}
   \caption{AUC change under standardization
   relative to Table~\ref{tab:auc}.}
   \label{tab:sensitivity-standardization}
   \centering
   \begin{tabular}{lcccc}
   \hline\hline
   Measure & GSE--Sgr & MW--GSE & MW--Sgr & Nissen \\
   \hline
   ASD  & 0.00 & 0.00 & 0.02 & 0.30 \\
   LSD  & 0.00 & 0.00 & 0.01 & 0.02 \\
   NLSD & $-$0.06 & 0.04 & 0.02 & 0.39 \\
   \hline
   \end{tabular}
   \tablefoot{Values are
   $\Delta\mathrm{AUC}=
   \mathrm{AUC}_{\rm scaled}-\mathrm{AUC}_{\rm raw}$ relative to Table~\ref{tab:auc}. Positive (negative) values indicate increased (decreased) separation after scaling. The affinity-kernel parameter is recalibrated to $\tau_{\rm scaled}=1.5$ to approximately preserve the raw relative weight spread $R(w)$.}
   \end{table}

   \begin{table}
   \caption{Mean AUC changes after leave-one-element-out relative to the seven-element analysis.}
   \label{tab:sensitivity-loo}
   \centering
   \begin{tabular}{lcccc}
   \hline\hline
   Measure & GSE--Sgr & MW--GSE & MW--Sgr & Nissen \\
   \hline
   ASD  &
   $0.00^{+0.00}_{-0.00}$ &
   $0.00^{+0.00}_{-0.00}$ &
   $0.02^{+0.15}_{-0.07}$ &
   $-0.05^{+0.07}_{-0.04}$ \\
   
   LSD  &
   $0.00^{+0.00}_{-0.00}$ &
   $0.00^{+0.00}_{-0.00}$ &
   $0.04^{+0.16}_{-0.07}$ &
   $-0.02^{+0.02}_{-0.01}$ \\
   
   NLSD &
   $-0.12^{+0.11}_{-0.10}$ &
   $-0.11^{+0.09}_{-0.05}$ &
   $0.00^{+0.05}_{-0.05}$ &
   $-0.02^{+0.04}_{-0.05}$ \\
   \hline
   \end{tabular}
   \tablefoot{Superscripts and subscripts give the distances from the mean to the maximum and minimum AUC changes, respectively, across removal of Mg, Al, Si, Ca, Ti, Cr, and Ni. All runs use the same seven-element complete-case star pools.}
   \end{table}
   \end{appendix}
   
\end{document}